# Accelerated spin-echo fMRI using Multisection Excitation by Simultaneous Spin-echo Interleaving (MESSI) with complex-encoded generalized SLIce Dithered Enhanced Resolution (cgSlider) Simultaneous Multi-Slice Echo-Planar Imaging

SoHyun Han[1,2], Congyu Liao[1,2], Mary Kate Manhard[1,2], Daniel Joseph Park[1], Berkin Bilgic[1,2], Merlin J. Fair[1,2], Fuyixue Wang[1,3], Anna I. Blazejewska[1,2], William A. Grissom[4], Jonathan R. Polimeni[1,2,5], Kawin Setsompop[1,2,5]

[1]*Athinoula A. Martinos Center for Biomedical Imaging, Massachusetts General Hospital, Charlestown, MA, USA*

[2]*Department of Radiology, Harvard Medical School, Boston, MA, USA*

[3]*Medical Engineering & Medical Physics, Harvard-MIT Division of Health Sciences and Technology, Cambridge, MA, USA*

[4]*Vanderbilt University Institute of Imaging Science, Vanderbilt University, Nashville, TN, USA*

[5]*Harvard-MIT Division of Health Sciences and Technology, Cambridge, MA, USA*

* Corresponding to: SoHyun Han, Ph.D.

Athinoula A. Martinos Center for Biomedical Imaging, Massachusetts General Hospital, Charlestown, MA, USA

Phone: 1-617-583-3245, Fax:1-617-726-7422

Email: hansomain256@gmail.com

**Abstract**

**Purpose:** Spin-echo functional MRI (SE-fMRI) has the potential to improve spatial specificity when compared to gradient-echo fMRI. However, high spatiotemporal resolution SE-fMRI with large slice-coverage is challenging as SE-fMRI requires a long echo time (TE) to generate blood oxygenation level-dependent (BOLD) contrast, leading to long repetition times (TR). The aim of this work is to develop an acquisition method that enhances the slice-coverage of SE-fMRI at high spatiotemporal resolution.

**Methods:** An acquisition scheme was developed entitled Multisection Excitation by Simultaneous Spin-echo Interleaving (MESSI) with complex-encoded generalized SLIce Dithered Enhanced Resolution (cgSlider). MESSI utilizes the dead-time during the long TE by interleaving the excitation and readout of two slices to enable 2× slice-acceleration, while cgSlider utilizes the stable temporal background phase in SE-fMRI to encode/decode two adjacent slices simultaneously with a 'phase-constrained' reconstruction method. The proposed cgSlider-MESSI was also combined with Simultaneous Multi-Slice (SMS) to achieve further slice-acceleration. This combined approach was used to achieve 1.5mm isotropic whole-brain SE-fMRI with a temporal resolution of 1.5s and was evaluated using sensory stimulation and breath-hold tasks at 3T.

**Results:** Compared to conventional SE-SMS, cgSlider-MESSI-SMS provides four-fold increase in slice-coverage for the same TR, with comparable temporal signal-to-noise ratio. Corresponding fMRI activation from cgSlider-MESSI-SMS for both fMRI tasks were consistent with those from conventional SE-SMS. Overall, cgSlider-MESSI-SMS achieved a 32× encoding-acceleration by combining $R_{\text{inplane}}$×MB×cgSlider×MESSI=4×2×2×2.

**Conclusion:** High-quality, high-resolution whole-brain SE-fMRI was acquired at a short TR using cgSlider-MESSI-SMS. This method should be beneficial for high spatiotemporal resolution SE-fMRI studies requiring whole-brain coverage.

## Introduction

Functional magnetic resonance imaging (fMRI) has been used as a powerful tool to investigate human brain function (1,2). It is well known that acquisition strategies can significantly affect the specificity and sensitivity of blood oxygenation level-dependent (BOLD) signals. Generally, gradient-echo (GE)-BOLD imaging is often used because of its ease of implementation and high contrast-to-noise ratio (CNR). However, GE-BOLD is highly sensitive to large draining veins (3–6) and suffers from signal dropout due to susceptibility effects near air-tissue interfaces (7). On the other hand, spin-echo (SE)-BOLD results in reduced CNR compared to GE-BOLD, which limits the use of SE-BOLD in fMRI studies at lower fields. SE-BOLD at 3T has similar intravascular (IV) and extravascular (EV) contributions, which reduces tissue sensitivity compared to GE-BOLD (8,9). Despite these disadvantages of SE-BOLD, studies have shown the advantage of SE-BOLD over GE-BOLD at 3T (8,10,11) in recovering the signal dropout near the regions of strong $B_0$ inhomogeneity.

At ultrahigh magnetic field strengths (e.g. 7 Tesla), SE-fMRI has also been shown to provide improved spatial specificity when compared to GE-fMRI, as it enhances the relative sensitivity of the BOLD signal from the parenchyma (12–14). Thus, SE-fMRI with enhanced spatial specificity can be useful to study brain organization and function at the cortical laminar or columnar levels (15–20). However, high spatiotemporal resolution SE-fMRI is difficult due to the long echo time (TE) needed to generate BOLD contrast (TE≈$T_2$ of gray matter (13,21)) and associated long repetition times (TR), along with higher specific absorption rate (SAR) from high flip-angle pulses. Nevertheless, achieving high spatiotemporal resolution as well as high spatial specificity is important in fMRI to investigate brain function at fine scales.

Although partial Fourier (22) and parallel imaging (23–25) techniques are very useful in reducing the number of phase-encoding steps in echo-planar imaging (EPI), these methods do not alleviate challenges in achieving whole-brain imaging with high spatiotemporal resolution. Recently, Simultaneous Multi-Slice (SMS) has been introduced to increase the temporal resolution of fMRI (26–30) while maintaining whole-brain slice-coverage, and the accelerated temporal sampling has been shown to be beneficial in several applications (31,32). Use of the CAIPIRINHA (27) (Controlled Aliasing In Parallel Imaging Results IN Higher Acceleration) technique in the form of blipped-CAIPI (30) for EPI can reduce the g-factor noise by shifting adjacent excited slices relative to each other in the phase-encoding direction and has been established as a standard technique in SMS-EPI (33). However, SE-SMS-EPI typically operates at low Multi-band (MB) factors (MB ≤ 3) due to peak power and SAR limitations, as well as $T_1$ saturation effects (34). Higher MB accelerations also introduce significant g-factor noise, especially when combined with in-plane acceleration (35).

Further slice-acceleration beyond conventional-SMS has been demonstrated with the Principles of Echo-Shifting with a Train of Observations (PRESTO) technique (31,32), which has been used for fMRI acquisitions (31,36–38). Other echo-shifting techniques (39–42) utilize the dead-time between excitation and readout, but these techniques are based on GE sequences. TE Interleaving imaging (43) and simultaneous echo refocusing (SER) (44) increase the number of slices per TR, up to three. However, to the best of our knowledge, echo-shifting techniques have not been combined with SE-EPI.

In this work, we introduce two complementary technologies (i) complex-encoded generalized SLIce Dithered Enhanced Resolution (cgSlider) and (ii) Multisection Excitation by Simultaneous Spin-echo Interleaving (MESSI) to achieve higher slice-accelerations in SE-fMRI. With cgSlider, temporally modulated RF-encodings between spatially adjacent simultaneously-acquired imaging sub-slices are used along with a phase-constrained reconstruction to achieve a 2× gain in slice-acceleration by taking advantage of the stable temporal background phase in SE. With MESSI, the dead-time during the long TE period in SE-fMRI (40,45) is used to interleave the excitation and readout of two imaging slices to provide an additional 2× slice-acceleration. cgSlider and MESSI can be combined, which can also be used in conjunction with conventional SMS parallel imaging. The 4× increase in slice-acceleration provided by cgSlider and MESSI does not come with additional g-factor penalty or any significant increase in peak RF power.

We demonstrate that cgSlider-MESSI-SMS enables whole-brain SE-fMRI acquisition at a nominal isotropic spatial resolution of 1.5mm, with a high temporal resolution of 1.5s and low image distortion and blurring ($R_{inplane}$=4). A total encoding-acceleration of 32× was achieved in this acquisition using $R_{inplane}$×MB×cgSlider×MESSI=4×2×2×2. SE-fMRI experiments at 3T using sensory stimulation and breath-hold tasks were used to demonstrate that the 4× enhancement in slice-coverage from cgSlider-MESSI-SMS can be achieved with minimal penalty when compared with conventional SE-SMS-EPI with the same temporal resolution.

## Theory

In this section, descriptions of the cgSlider and MESSI techniques are provided. Each of these methods can achieve a two-fold slice-acceleration and can be used jointly, along with conventional-SMS acceleration, to achieve high slice-accelerations in SE-fMRI.

### *Complex-encoded gSlider acquisition (cgSlider)*

Two adjacent sub-slices are acquired together using complex-encoded gSlider RF-encoding, where the excitation phase of one of the sub-slices (blue colored sub-slice in Fig. 1A) is modulated across the time frames, as shown in Fig. 1A, which can be described as:

$$S_{\text{cg}}(n) = S_{\text{A}}(n)e^{j\phi_{\text{A}}(n)} + S_{\text{B}}(n)e^{j\phi_{\text{B}}(n)}e^{j\theta(n)}, \begin{cases} \theta(n) = \dfrac{\pi}{2} \text{ if } n \text{ is odd} \\ \theta(n) = -\dfrac{\pi}{2} \text{ if } n \text{ is even} \end{cases} \quad (1)$$

where $S_{\text{cg}}$ is the cgSlider signal acquired at each time point ($n$) consisting of the combination of two simultaneously-encoded adjacent sub-slices. $S_{\text{A}}$ and $S_{\text{B}}$ are the magnitudes of sub-slices A and B, $\phi_{\text{A}}$ and $\phi_{\text{B}}$ are the corresponding background phases of sub-slices A and B, respectively. Here, $\theta(n)$ denotes the temporally-modulated RF-encoding phase of sub-slice B at the $nth$ temporal frame. A Shinnar-Le Roux (SLR) (46) based cgSlider RF pulse (34) was used for the 90° excitation pulse with a time-bandwidth-product (TBWP) of 9, in conjunction with a standard SLR pulse for the 180° refocusing with a TBWP of 5. This design provides no increase in the 180° peak voltage and approximately the same 90° peak-voltage when compared to standard single-slice acquisition. It is important to note the use of complex-valued signal modulation of sub-slice B and the addition of a time-varying phase modulation, which enables separation of the two sub-slices from the acquired slab-signal using the reconstruction method described below. Note that the division of the slab into sub-slices will result in different inflow effects for each of the two sub-slices due to inflowing spins from above and below the slab. In this initial work we will not investigate these effects but plan to investigate them more quantitatively in future work.

### cgSlider image reconstruction

*'Sliding-window' reconstruction:* The complex-encoding described above enables conventional Hadamard encoding reconstruction methods (47) across two adjacent time frames. For example, with the 1st and 2nd time frames, it is assumed that the underlying sub-slice image magnitudes and phases $S_A$, $S_B$, $\phi_A$, and $\phi_B$ are slowly varying between these two adjacent time frames in the SE-fMRI acquisition $(\phi_A(1) \approx \phi_A(2), \phi_B(1) \approx \phi_B(2), S_A(1) \approx S_A(2), S_B(1) \approx S_B(2))$. In other words, due to the refocusing of a spin-echo, we assume that both the magnitude of the signal and, more critically, the phase of the signal are not changing in the brain over this short time frame—that is, physiological processes such as those driven by the cardiac or respiratory cycles as well as neuronal activity are assumed to not cause a substantial change in the image phase. Under this assumption, Eqn. (1) can be rewritten as:

$$\begin{cases} S_{cg}(1) = S_A(1)e^{j\phi_A(1)} + jS_B(1)e^{j\phi_B(1)} \\ S_{cg}(2) = S_A(2)e^{j\phi_A(2)} - jS_B(2)e^{j\phi_B(2)} \end{cases} \quad (2)$$

The signal magnitude and phase of each sub-slice can then be obtained by adding or subtracting the cgSlider sub-slices signal of two adjacent time frames, which can be written as:

$$\begin{cases} S_{cg}(1) + S_{cg}(2) \approx 2S_A(1.5)e^{j\phi_A(1.5)} \\ S_{cg}(1) - S_{cg}(2) \approx j2S_B(1.5)e^{j\phi_B(1.5)} \end{cases} \quad (3)$$

where the signals from the two sub-slices at the time point half-way between the two acquisitions (i.e., at $n = 1.5$) are expressed as linear combinations of the slab signals measured at time point 1 and time point 2. This expression can be applied to subsequent time points to provide a reconstruction of the sub-slice time-series data by a 'sliding-window' reconstruction in which each reconstructed time-point results from the linear combination of the two surrounding time points, as illustrated in Fig. 1B with black arrows. This method allows simple separation of the cgSlider signal while obtaining an image SNR benefit by a factor of $\sqrt{2}$ due to noise averaging. However, this method also causes temporal blurring effects owing to the sharing of data across two adjacent time points.

*'Phase-constrained' reconstruction*: The sliding-window reconstruction makes a strong assumption about both the magnitude and phase of the acquired image data being slowly varying over time such that neither components of the complex-valued signal change appreciably from one time point to the next. However, this strong assumption can be relaxed for the signal magnitude. It is observed that the signal change in SE-fMRI is mostly confined to the image magnitude, and there is little change in the background phase (shown in Supporting Information Figure S1). By taking advantage of the observed negligible temporal phase variations in SE-fMRI, the background phase reconstructed from the sliding window method ($\phi_{\text{A-SW}}$ and $\phi_{\text{B-SW}}$ in Fig. 1B) can be used as an initialization for a phase-constrained reconstruction so that the number of unknown values in Eqn. (1) is reduced from 4 to 2; in this approach the sliding-window reconstruction provides a reference phase to enable the reconstruction of the magnitude of the two imaging sub-slices directly from each acquired time frame without the temporal blurring induced by the sliding-window approach. However, the time point of the estimated phase from the sliding-window approach is $\phi_{\text{A-SW}}(n+0.5)$ or $\phi_{\text{B-SW}}(n+0.5)$. To match the number of time points, $n'$ was set as $(n-0.5)$ and the last time point was repeated with $(last\ time\ point - 0.5)$. Here, the phase-constrained reconstruction estimates the sub-slice signals $S_{\text{A-PC}}(n)$ and $S_{\text{B-PC}}(n)$ by solving the following linear system of equations through simple matrix inversion:

If $n$ is odd,

$$\begin{bmatrix} \text{Re}(S_{\text{cg}}(n)e^{-j\phi_{\text{A-sw}(n')}}) \\ \text{Im}(S_{\text{cg}}(n)e^{-j\phi_{\text{A-sw}(n')}}) \end{bmatrix} = \begin{bmatrix} 1 & -\sin\Delta\phi_{\text{SW}}(n') \\ 0 & \cos\Delta\phi_{\text{SW}}(n') \end{bmatrix} \begin{bmatrix} S_{A-\text{PC}}(n) \\ S_{B-\text{PC}}(n) \end{bmatrix} \quad (4)$$

If $n$ is even,

$$\begin{bmatrix} \text{Re}(S_{\text{cg}}(n)e^{-j\phi_{\text{A-sw}(n')}}) \\ \text{Im}(S_{\text{cg}}(n)e^{-j\phi_{\text{A-sw}(n')}}) \end{bmatrix} = \begin{bmatrix} 1 & \sin\Delta\phi_{\text{SW}}(n') \\ 0 & -\cos\Delta\phi_{\text{SW}}(n') \end{bmatrix} \begin{bmatrix} S_{A-\text{PC}}(n) \\ S_{B-\text{PC}}(n) \end{bmatrix} \quad (5)$$

where $\Delta\phi_{\text{SW}}$ is the phase difference calculated between the two sub-slices from the sliding-window reconstruction, i.e., $\Delta\phi_{\text{SW}} = \phi_{\text{B-SW}} - \phi_{\text{A-SW}}$. Eqn. (4) and Eqn. (5) are derived in Supporting background information.

### Multisection Excitation by Simultaneous Spin-echo Interleaving (MESSI) sequence

A schematic diagram of the MESSI pulse sequence is shown in Fig. 2. To acquire two imaging slices jointly in an interleaved fashion (denoted as MESSI-1 and MESSI-2), the following four sequence components were added to conventional SE-EPI sequence. First, an additional readout and 90° and 180° pulses for the MESSI-2 slice (blue-colored RF pulses and readout) with a TE matched to that of the MESSI-1 slice (red-colored RF pulses and readout) were added. Second, to separate the *k*-space signals of the two MESSI slices, dephasing gradients (green-colored gradients) that shift the signal of the different slices in-plane were added before the 180° pulse of the MESSI-2 slice. Gradient moment parameters α and β correspond to $k_{max}/2$ of frequency encoding and phase encoding, respectively, and $k_{factor}$ is the integer-valued scaling parameter determining the distance in *k*-space between the two MESSI slices. As $k_{factor}$ is increased by one, the distance between the signal of the two slices is increased by $k_{max}$ in frequency and phase encoding directions, which acts to prevent *k*-space signal leakage between slices. Third, prior to the data acquisition of the MESSI-1 slice, rephasing gradients (red-striped gradients) were inserted to rephase the signal for MESSI-1 slice. During data acquisition of MESSI-1 slice, the spins from MESSI-2 slice are dephased. For the same reason, rephasing gradients for MESSI-2 slice (blue-striped gradients) were inserted. Fourth, to avoid free induction decay (FID) signal from 180° RF pulse of MESSI-2 slice introduced by imperfect RF refocusing pulse, spoiler gradients were added (purple-striped gradients).

*$k_{factor}$ optimization*: The effect of the inserted MESSI echo-shifting (dephasing and rephasing) gradients on the spins from the two MESSI-slices in the cases where $k_{factor}$ is set to 1 or 2 is illustrated in Figs. 3A and 3B, respectively. Rephasing gradients for MESSI-1 slice (red-striped gradients) rephase spins in the readout for MESSI-1, while dephasing magnetization from MESSI-2 slice. The same is true for the rephasing gradients (blue-striped gradients) for MESSI-2 slice. Increasing the value of $k_{factor}$ increases the signal dephasing between MESSI slice groups and reduces the potential for signal leakage between the slices for data at the edges of *k*-space.

### cgSlider-MESSI-SMS sequence

Fig. 4 describes how the cgSlider, MESSI and conventional SMS techniques can be combined synergistically to provide high slice-accelerations in SE-fMRI. MESSI enables an increase in the slice-coverage by exploiting dead-time and exciting an additional slice group (red and blue slices) per TR, whereas cgSlider allows an increase in coverage by exciting complex-encoded spatially-adjacent sub-slices (red slab). Combining these techniques (cgSlider-MESSI) enables simultaneous excitation of additional slice groups and their complex-encoded spatially-adjacent slices (red and blue slabs). Inclusion of SMS further extends the slice coverage through exciting cgSlider-MESSI-1 and cgSlider-MESSI-2 groups (yellow and green slabs), spaced apart evenly across the FOV in the $z$ direction.

### *Velocity-encoding phase correction and reference phase acquisition*

There are three phase components in the image produced by cgSlider-MESSI-SMS: the background phase, the velocity-encoding phase from additional gradients for the MESSI sequence implementation, and the change in phase due to the fMRI activation. Large dephasing/rephasing gradients in the MESSI sequence can introduce non-negligible velocity-encoding, which can induce phase variations due to respiration/cardiac induced movement and head motion (48–51). Such phase variations can affect the cgSlider-MESSI reconstruction, causing striping artifacts in the reconstructed images along the slice-direction. An approach to remove this phase corruption was developed that takes advantage of the fact that this phase corruption is typically spatially smooth and should not vary substantially across the thin slab of the cgSlider-encoding. In this approach, first, the phase images (denoted as $\angle(S_{\mathrm{cg}}(\mathrm{t}))$) were averaged separately for all odd- and even-numbered frames of the time series data of the cgSlider slab-encoded signal as shown in Fig. 5 in the green box ($\mathrm{avg}\angle(S_{\mathrm{cg}})$). Second, the phase difference between each time frame and the averaged phase ($\angle(\mathrm{difference}) = \angle S_{\mathrm{cg}}(\mathrm{t}) - \mathrm{avg}\angle(S_{\mathrm{cg}})$) was calculated separately for the odd and even time frames. Third, because of the background phase is well known to be smoothly varying in SE images, a spatial filter was applied to the phase difference image ($\angle(\mathrm{difference\_filtered})$), to reduce noise and more accurately estimate the velocity-encoding phase variation, and the estimated velocity-encoding phase was removed, leaving behind the background phase and the phase change related to fMRI activation. Finally, the phase-constrained reconstruction was performed after this velocity-encoding phase correction. However, this velocity phase removal process is not perfect, which can remain striping artifacts. An alternative approach was also examined, neglects the temporal phase changes related to fMRI activation, which should be relatively small. With this assumption, the phase of the cgSlider-MESSI-SMS was replaced by a reference phase from a cgSlider-SMS with matching sequence parameters that contains only the background phase with no velocity-encoding phase contamination.

## Methods

### Participants

Nine healthy subjects (5 male, 4 female), aged 25–39 years old, participated in this study. All procedures followed the guidelines of the Institutional Review Board of the Massachusetts General Hospital and Sungkyunkwan University. Procedures were fully explained to all subjects, and informed written consent was obtained before scanning in accordance with the Declaration of Helsinki.

### MRI Acquisition

All measurements were performed on a 3T scanner (MAGNETOM Prisma, Siemens Healthineers, Erlangen, Germany) with the vendor-supplied 32-channel head coil and the vendor supplied 64-channel head and neck coil. The developed sequence was combined with the blipped-CAIPI SMS technique (30) at MB=2 to further increase slice-coverage and capture the entire brain in a single repetition, and $R_{inplane}$=4 was used to minimize image distortion and blurring. VERSE (52) was also applied to the MB cgSlider RF pulses to reduce peak voltage and SAR. VERSE was applied to both 90° and 180° pulses to ensure that the slice profile degradations and shifts at off-resonance are similar across excitation and refocusing to achieve good signal level (34). The sequence parameters used here are as follows: TR/TE = 1500/75 ms, $FOV_{xy}$ = 210 × 210 mm$^2$, partial Fourier = 6/8, 1.5 mm isotropic resolution, effective echo spacing (ESP) = 0.173 ms. The readout bandwidth parameter value was chosen here to minimize the echo spacing in order to minimize EPI blurring and distortion, as is commonly done for conventional EPI. The reference phase acquisition was collected using cgSlider-SMS with a TR of 2500 ms; this longer TR was required to match the number of slices between the reference data and the accelerated cgSlider-MESSI-SMS data. The reference data were acquired at the beginning of each run prior to fMRI data collection and therefore introduced a small increase in total scan duration (2.5 s per run).

### $k_{factor}$ optimization in MESSI sequence

To examine the level of signal leakage between MESSI slices, direct measurements of the signal leakage levels were obtained in the MESSI sequence by setting either MESSI-1 or MESSI-2 RF excitation pulse flip angles to 0° for acquisitions. Two $k_{factor}$ settings of 1 or 2 were evaluated.

***Velocity-encoding phase correction in both cgSlider and cgSlider-MESSI***

For both cgSlider and cgSlider-MESSI cases, reconstructions were performed with and without velocity-encoding phase correction to assess tSNR level improvement (see below for description of tSNR comparisons).

***tSNR comparisons***

For tSNR analysis, three protocols were compared with four subjects: conventional-SMS, cgSlider-SMS, and cgSlider-MESSI-SMS. To achieve an unbiased comparison, the MR parameters and TR were kept constant and the total number of slices were adjusted accordingly to the net slice-acceleration factor of each protocol. In summary, we compared (i) conventional-SMS with $R_{\text{inplane}} \times MB = 4 \times 2$, 26 slices (100% slice gap), $FOV_z = 78mm$, (ii) cgSlider-SMS with $R_{\text{inplane}} \times MB \times cgSlider = 4 \times 2 \times 2$, 52 slices (no slice gap), $FOV_z = 78mm$, and (iii) cgSlider-MESSI-SMS with $R_{\text{inplane}} \times MB \times cgSlider \times MESSI = 4 \times 2 \times 2 \times 2$, 84 slices (no slice gap), $FOV_z = 126mm$ (whole-brain coverage), $k_{\text{factor}} = 2$. The number of repetitions (NR) was 140 for each protocol, corresponding to a total acquisition time of 3mins 30secs per protocol. The tSNR maps were calculated from 100 NRs, excluding 20 NRs at both the beginning and at the end, by dividing the temporal mean of the time series by the temporal standard deviation. Additionally, average and standard deviation of the resulting tSNR were calculated in ROIs defined as brain in four subjects.

***Visual/breath-hold fMRI activation***

To assess the performance of cgSlider-MESSI-SMS compared to conventional SMS in SE-fMRI, fMRI data were acquired using the three protocols described above: conventional-SMS, cgSlider-SMS, and cgSlider-MESSI-SMS. For the visual stimulation session, three subjects were presented with a standard flashing scaled-checkboard stimulus (12s on, 20s off, 4 on-off blocks per run); each run lasted 210s, and four runs were acquired for each protocol that were averaged together during the analysis. For the timed breath-hold task, the subject was cued to hold their breath for 12s followed by 30s of free breathing with four-breath-holds/run, and seven runs were acquired for each protocol averaged during analysis. FSL (http://www.fmrib.ox.ac.uk/fsl) was used to perform fMRI analysis; spatial smoothing (3mm kernel) was applied for the breath-hold task to boost SNR but not for the visual stimulation task where SNR is sufficient, while MCFLIRT motion correction was applied to both.

## Results

Supporting Information Figure. S1A shows the z-statistic maps for a visual stimulation task obtained using the time-series image magnitude and the time-series image phase of a single conventional-SMS SE-fMRI acquisition. The results demonstrate that the BOLD responses in conventional-SMS SE-fMRI is mostly confined to the image magnitude, and little/no change in the corresponding background phase was detected in response to activation. Supporting Information Figure. S1B shows the estimated background phases for two adjacent sub-slices from the cgSlider acquisition. The phases of the adjacent sub-slices were similar to each other, which supports the feasibility of the proposed phase-constrained reconstruction approach.

To quantify signal leakage as a function of the value of $k_{factor}$ we acquired test data in one subject with different parameter values. Fig. 6 shows the results of this analysis, including the reconstructed images and the signal leakage maps corresponding to signal from one slice in the MESSI slice group leaking into the other slice. Figs. 6A and 6B show the signal leakage between the MESSI slice groups for acquisitions with $k_{factor}$ of 1 and 2. Upper and lower rows show the results from the cases, when MESSI-1 or MESSI-2 pulses were set to 0°. With $k_{factor}=2$, the signal leakage between MESSI groups is negligible, whereas with $k_{factor}=1$, the leakage is clearly seen. The leakage maps were all multiplied by a factor of 10 relative to the brain images to visualize the leakage pattern. In order to avoid signal leakage, a $k_{factor}$ of 2 was used for MESSI acquisitions.

The left column of Fig. 7A shows tSNR maps without velocity-encoding phase correction, at varying $k_{factor}$ from 1 to 4 to evaluate the effect of large dephasing/rephasing gradients in the MESSI sequence. Higher $k_{factor}$ results in lower tSNR, which reflects that larger gradients induce higher sensitivity to potential image phase variations due to respiration/cardiac induced movement and head motion. To overcome tSNR deterioration, velocity-encoding phase correction was applied and showed comparable tSNR level despite the stronger dephasing/rephasing gradients from the increased $k_{factor}$, as shown in right column of Fig. 7A. Also, a comparison of the reconstruction without correction, with the velocity-encoding phase correction, and with reference phase is shown in Fig. 7B, where sagittal reformats of axially-acquired slices are presented. The striping artifact across slices, was substantially reduced by the velocity-encoding phase correction, but not perfectly removed. For example, the white arrow points to an area that shows the reduced striping artifact both with velocity-encoding phase correction and with reference phase. However, the yellow arrow points to an area showed less striping with the reference phase than with the velocity-encoding phase correction, which reflects that the striping artifact is mostly originated from the velocity-encoding due to the additional gradient lobes in the MESSI sequence.

The reconstructed conventional-SMS, cgSlider-SMS, and cgSlider-MESSI-SMS images were compared in terms of overall image quality as well as the resulting tSNR before and after velocity-encoding phase correction (see Fig. 8). The cgSlider-SMS and cgSlider-MESSI-SMS tSNR maps, before and after velocity-encoding phase correction, are shown in Figs. 8B and 8C. There is no apparent difference in tSNR maps among conventional SMS, cgSlider-SMS, and cgSlider-MESSI-SMS with velocity-encoding phase correction. However, although the time bandwidth product of RF pulses in conventional-SMS were matched with that of cgSlider-SMS, slightly higher tSNR values were seen for the cgSlider-SMS reconstruction when compared to that from conventional-SMS. This is likely caused by the expected small increase in signal level in cgSlider-SMS due to improved signal refocusing performance. In particular, the refocusing pulse for the cgSlider-SMS acquisition extends across the two sub-slices of cgSlider, with each sub-slice experiencing only one transition band with incomplete refocusing, rather than two in the conventional-SMS. For further assessment, average and standard deviations of tSNR obtained from four subjects were compared. Average values±standard deviations for conventional-SMS, cgSlider-SMS, and cgSlider-MESSI-SMS were 9.2±1.0, 10.6±1.4, and 10.6±1.2, respectively. Average tSNR values among different methods were not statistically significant. Our velocity-encoding phase correction resulted in comparable tSNR to the conventional-SMS (Fig. 8C). However, tSNR maps from cgSlider-MESSI-SMS without velocity-encoding phase correction showed much lower tSNR than other protocols (Fig. 8B) due to the increased velocity-encoding induced by the dephasing/rephasing gradients used for MESSI.

In particular, the proposed cgSlider-MESSI-SMS approach achieves whole-brain coverage at 1.5mm isotropic resolution with 1.5s temporal resolution. In the presented conventional-SMS there is a 100% gap imposed to allow for this acquisition, to have the same brain-coverage as cgSlider-SMS, albeit with half the number of reconstructed slices.

Finally, Fig. 9 demonstrates the feasibility of SE-fMRI using cgSlider-MESSI-SMS by comparing the resulting BOLD activation maps with those values from conventional-SMS and cgSlider-SMS. For each acquisition, z-statistic maps (thresholded at $p < 0.01$) are overlaid on a single reconstructed image of the corresponding acquisition. Fig. 9A shows z-statistical maps from two adjacent-slices, demonstrating the feasibility of the proposed cgSlider reconstruction to enable both anatomical details and functional activation patterns from two adjacent sub-slices. For both the visual stimulation and breath-hold fMRI experiments, cgSlider-MESSI-SMS maintains the same temporal resolution at 2× brain coverage when compared to that from cgSlider-SMS, while exhibiting comparable activation patterns. The similarity of activation maps between these acquisitions indicates that fMRI sensitivity is not compromised with the addition of MESSI. For further validation, z-statistic map comparisons among different methods were shown with two subjects in Supporting Information Figure S2, also mean and standard deviation of thresholded z values ($z > 2$), and the number of activated voxels (clusters of minimum size of 30 voxels) were calculated in Supporting Information Table S1. Especially in Fig. 9B, activations were detected in the medial prefrontal cortex (yellow arrow) are known to be nearby regions with large susceptibility gradients, which are difficult to detect with GE-EPI sequence (8,11).

## Discussion

Here we proposed a new method for accelerating SE-fMRI using the cgSlider-MESSI-SMS acquisition. The cgSlider component enables a two-fold increase in slice-acceleration using a phase-constrained reconstruction that utilizes the spatiotemporal smoothness of SE-image phase. The MESSI component provides a ~2× higher efficiency in slice acquisition by interleaving excitation and data collection within the sequence dead-time, taking advantage of the long TE of BOLD-weighted SE-fMRI. Finally, conventional-SMS was further combined with both cgSlider and MESSI, resulting in a total ~32× acceleration factor ($R_{inplane}$ ×MB×cgSlider×MESSI=4×2×2×2). Our cgSlider-MESSI-SMS approach successfully demonstrated 1.5mm isotropic whole-brain coverage at a temporal resolution of 1.5s, which is not feasible with conventional SE-SMS-EPI. The feasibility of SE-fMRI with this new sequence was demonstrated for both sensory stimulation and breath-hold tasks in healthy subjects at 3T, which showed comparable z-statistic maps to those from conventional SE-SMS-EPI at the same temporal resolution but at four times greater slice-coverage. Large slice-coverage with high spatiotemporal resolution should be particularly useful in resting-state fMRI studies, where whole-brain acquisitions are required (53).

When the echo-shift method is applied to SE-EPI, the target TE will limit the readout window. In this work, we have achieved echo-shifting factor of 2 by applying in-plane acceleration factor of 4, allowing us to shorten the readout window. With our current scheme, a higher echo-shifting factor is not achievable without a significant TE increase. The MB-factor was limited to 2, in order to limit the total acceleration factor to 8 when using a 32-channel coil array at 3T (we used $R_{\text{inplane}} \times \text{MB} = 4 \times 2$) (54). At higher field strength, acceleration performance increases and so a higher acceleration factor can be achieved (55), which would benefit the reduction of readout window further.

It is worthwhile to note that while the use of higher $k_{\text{factor}}$ and/or higher spatial resolution acquisition can reduce the leakage of high frequency signal between the MESSI data groups as shown in Fig. 6, the velocity encoding and hence the image phase corruption would also increase due to the stronger gradients, as demonstrated in Fig. 7A. Moreover, a higher $k_{\text{factor}}$ results in a longer TE, lower SNR, and additional diffusion effect, which is not desirable for fMRI applications. Due to these potential artifacts from increasing $k_{\text{factor}}$, it is necessary to optimize $k_{\text{factor}}$ to minimize the strength of gradients while still avoiding leakage of high frequency signal between the MESSI groups. The additional diffusion effect of cgSlider-MESSI-SMS sequence with $k_{\text{factor}} = 2$ was calculated, and it was found that the maximum b-value was 1.28 s/mm$^2$, which should not significantly affect the fMRI signal. The detrimental effect of this on the cgSlider reconstruction can be mitigated by the velocity-encoding phase correction step, as shown in Fig. 7B. With this correction, cgSlider-MESSI-SMS results in nearly the same tSNR as that of cgSlider-SMS while providing increased slice-acceleration as shown in Fig. 8. However, there are remaining stripes across slices, as shown in Fig. 7B, implying that velocity/motion artifacts were not perfectly corrected even though the velocity-encoding phase correction provides improved image quality. As an alternative approach in correcting phase corruption from velocity-encoding, a reference phase that utilizes a cgSlider-SMS 'pre-scan' is proposed and shown to significantly reduce the striping artifacts. Moreover, the fMRI z-statistic maps from the reconstruction using this reference phase were compared to those using the velocity-encoding phase correction in Supporting Information Figure S2. While the reference phase does not incorporate the phase changes due to the fMRI activation into the reconstruction, the estimated fMRI activation from such approach was found to be comparable to that of the velocity-encoding phase correction approach, suggesting that the phase changes due to the fMRI activation are small relative to the velocity-encoded phase variation.

Based on the timing of the sequence, the target BOLD-weighted protocol can achieve a minimal volume acquisition time of 1.2s while maintaining whole brain coverage with 1.5mm isotropic resolution, however, the current implementation is limited by SAR. Typical SAR level in the cgSlider-MESSI-SMS approach was ~95% of the 6-min SAR limits. However, the use of lower flip angles to avoid overflipping in the center of the brain, such as 78° for excitation and 160° for refocusing (56), reduce RF power deposition and thus could also be explored with cgSlider-MESSI-SMS. This should reduce SAR level to ~75.5%. On the other hand, when the TR value is short compared to the tissue $T_1$ value (e.g. TR = 1.5s), the Ernst angle will maximize signal, however the Ernst angle for the excitation pulse of a spin-echo acquisition is typically larger than 90°. If we consider the SAR resulting from this higher excitation flip angle, which here would be ~113° at 3T, the SAR level would increase to ~103.8%. Therefore, the optimization of RF flip angles must balance between maximizing signal levels and achieving signal uniformity over the tissue of interest while remaining within the SAR limits. Even though the peak power of the RF pulses in the sequence does not increase when cgSlider and MESSI are employed, the total SAR increases by ~4× from the 4× increase in the number of slices being excited and refocused. In this work, the VERSE algorithm was applied to the RF pulses to help reduce SAR, which can result in some compromise in image quality in regions of strong $B_0$ inhomogeneity. Future work will explore the use of alternative pulse design approaches (57) and parallel transmission to help reduce SAR (58,59). In particular, Power Independent Number of Slices (PINS) pulses (60–63) can be used to reduce SAR at the high MB factor.

Future work will also focus on the application of this method to ultra-high field SE-fMRI at 7T, where the $T_2$ weighting can provide enhanced microvascular specificity and higher spatial resolution imaging. At ultra-high fields, the optimal TE value for SE-BOLD are shorter, which requires high in-plane acceleration. However, a higher in-plane acceleration factor might be acceptable because of reduced g-factor penalties (55) at 7T. A multi-shot approach would allow for reduced echo-train-lengths, at the cost of advanced reconstruction techniques to overcome artifacts due to phase variations across shots (64,65). A reduced FOV acquisition with outer volume suppression (66,67) will also be explored to further reduce echo-train-lengths, which in theory should not greatly affect image quality.

Given the parameters in this work, the main source of BOLD signal would be thermal noise. The tSNR maps shown in Fig 8B reflects this, where the highest tSNR values are on the outer part of the brain closest to the receiver coil, reflecting a thermal noise dominated acquisition (68).

## Conclusion

Here we proposed a new method, cgSlider-MESSI-SMS, and demonstrated that this can provide whole-brain SE-fMRI acquisitions at 3T with a spatial resolution of 1.5mm and temporal resolution of 1.5s through achieving a total acceleration factor of 32-fold using $R_{\text{inplane}} \times MB \times cgSlider \times MESSI = 4 \times 2 \times 2 \times 2$. With this newly developed pulse sequence and associated image reconstruction approaches, SE-fMRI experiments at 3T using sensory stimulation and breath-hold tasks successfully demonstrated the $4\times$ enhanced slice-coverage with minimal SNR penalties.

## Acknowledgements

This work was supported in part by NIH grants R01-EB020613, R01-EB019437, R01-MH116173, R01-MH111419, U01-EB025162, P41-EB015896, by the MGH/HST Athinoula A. Martinos Center for Biomedical Imaging, and by IBS-R015-D1; and was made possible by the resources provided by NIH Shared Instrumentation Grants S10-RR023401, S10-RR019307, S10-RR019254 and S10-RR023043.

**Figure captions**

**Figure 1.** **(A)** Illustration of complex-valued gSlider RF-encoding. Temporally varying phase modulation of $\pm \pi/2$ was applied to the second sub-slice (blue text). **(B)** Illustration of the 'sliding-window' reconstruction. By assuming signal magnitudes and background phases are slowly varying between time frames, signal magnitudes and background phases for each slice were estimated through a 'sliding-window' reconstruction shown as black arrows. **(C)** After estimating background phases from the sliding-window reconstruction, the 'phase-constrained' reconstruction estimates signal magnitudes for sub-slices A and B at each time frame directly without temporal smoothing

**Figure 2.** MESSI sequence diagram, showing two interleaved slices (MESSI-1, red and MESSI-2, blue). The dephasing gradients separate the $k$-spaces of the two MESSI groups (green-striped gradients). $\alpha$ and $\beta$ correspond to $k_{max} / 2$ values along the readout and phase encoding directions, respectively. $k_{factor}$ is the integer-valued scaling parameter that determines the distance of the $k$-space centers between the two MESSI groups. The rephasing gradients are for ensuring that the $0^{th}$ moments of all gradients are zero prior to acquiring the MESSI-1 or MESSI-2 readouts (red- and blue- striped gradients). Spoiling gradients (purple-striped gradients) were added to avoid possible artifacts from FID signals arising from imperfect RF refocusing pulses.

**Figure 3.** **(A)** and **(B)** represent the MESSI sequence diagram and phase evolutions along the readout gradient for spins from the two MESSI groups, for the cases where the integer-valued $k_{factor}$ parameter is set to 1 or 2, respectively. The red solid line and blue dashed line correspond to the MESSI-1 and MESSI-2 groups, respectively. As $k_{factor}$ is increased, the distance between two $k$-spaces of the two MESSI groups is increased, as shown with blue and red boxes. Numbers listed within the gradient lobes signify the relative value of the $0^{th}$ moment.

**Figure 4.** Overview of the combination of cgSlider, MESSI and conventional SMS techniques. MESSI enables an increase in the slice coverage by exciting an additional slice group (red and blue slices), whereas cgSlider allows an increase in coverage by exciting complex-encoded spatially-adjacent slices (red slab). Combining the techniques (cgSlider-MESSI) enables simultaneous excitation of additional slice groups and their complex-encoded spatially-adjacent

slices (red and blue slabs). Inclusion of SMS extends the slice coverage through exciting cgSlider-MESSI-1 and cgSlider-MESSI-2 groups (yellow and green slabs), spaced apart evenly across the $FOV_z$.

**Figure 5.** Overview of the phase correction process for cgSlider and cgSlider-MESSI acquisitions: First, the averaged phases for odd- and even- numbered time frames were calculated (green box). The phase difference between each time frame and averaged phases ($\angle$(difference)) was found, before a Hamming filter was then applied $\angle$(difference_filtered) to subtract the velocity encoding phase that is sensitive to physiological changes such as motion. After the velocity encoding phase correction, the 'phase-constrained' reconstruction was performed.

**Figure 6.** Characterization of potential $k$-space signal leakage expected in the high spatial frequencies, as a function of the $k_{factor}$ parameter value. To directly investigate the level of signal leakage between two MESSI groups, either MESSI-1 or MESSI-2 excitation pulses were set to 0°. **(A)** and **(B)** display maps of signal leakage between MESSI groups for the case of $k_{factor} = 1$ and 2, respectively. Red and blue tinted frames represent MESSI-1 and 2 groups, respectively. With $k_{factor} = 2$ the signal leakage between MESSI groups is negligible, whereas with $k_{factor} = 1$ leakage is clearly seen (white arrows). The leakage maps were all multiplied by a factor of 10 relative to the brain images to visualize the leakage pattern.

**Figure 7.** **(A)** tSNR maps without velocity encoding phase correction, varying $k_{factor}$ from 1 to 4 to evaluate the effect of large dephasing and rephasing gradients in the MESSI sequence (left column). Corresponding tSNR maps after the application of velocity encoding phase correction in the reconstruction, at varying $k_{factor}$ from 1 to 4 (right column). **(B)** Reconstructed cgSlider-MESSI-SMS sagittal images with no correction (top), with velocity encoding phase correction (middle), and with reference phase (bottom).

**Figure 8.** Comparisons of image quality **(A)** and corresponding tSNR before **(B)** and after **(C)** velocity encoding phase correction for conventional SE SMS, cgSlider-SMS, cgSlider-MESSI-SMS. Similar image quality and tSNR levels were achieved among different methods with velocity encoding phase correction. For both cgSlider-SMS and cgSlider-MESSI-SMS, phase correction resulted in an improved tSNR. The improvement is particularly significant in cgSlider-MESSI-SMS, which contains more shot-to-shot image phase variations due to the increased velocity

encoding induced by the echo shifting (dephasing and rephrasing) gradients. Intensity correction was performed for (A).

**Figure 9.** Functional activation maps, represented as z-statistics of the detected BOLD response, resulting from visual stimulation and breath-hold fMRI datasets (spatial smoothing was applied for the breath-hold task to boost SNR but not for the visual stimulation task where SNR was sufficient). The cgSlider-MESSI-SMS method maintains high-temporal resolution at 4×, and 2× brain coverage compared with conventional SE-SMS and cgSlider-SMS. Also, the cgSlider-MESSI-SMS achieves a comparable extent of activation compared to cgSlider-SMS for both the visual stimulation **(A)** and breath-hold task **(B)**. Intensity correction was performed for all images.

**Supporting Information Figure S1. (A)** Z-statistical maps obtained from signal intensity (top) and phase data (bottom) from a visual stimulation task. Detected fMRI responses are found only in the signal magnitude reconstruction with little/no fMRI responses detected in the phase image. **(B)** Estimated background phase from cgSlider acquisition and sliding-window reconstruction for two adjacent slices (A, B sub-slices depicted in Fig.1.) and three time points (1.5, 2.5, 3.5) with little phase variation over time and between sub-slices.

**Supporting Information Figure S2.** Z-statistic maps from visual stimulation with **(A)** conventional SMS, **(B)** cgSlider-SMS, **(C)** cgSlider-MESSI-SMS after velocity encoding phase correction, and **(D)** cgSlider-MESSI-SMS after correction using the reference phase. Intensity correction was performed for all images.

# Figures

## Figure 1.

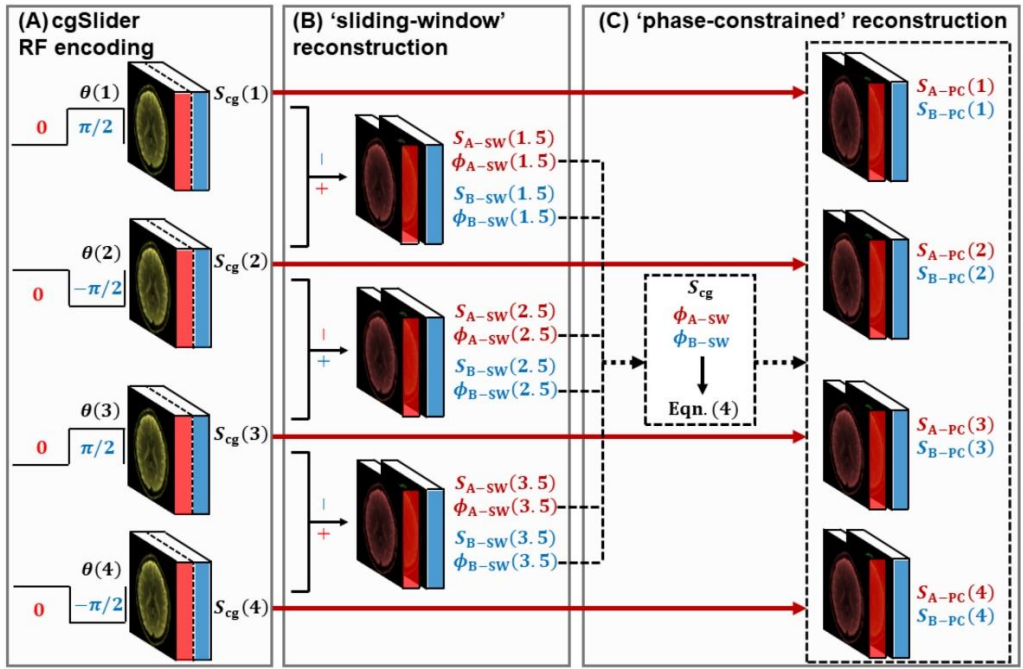

## Figure 2.

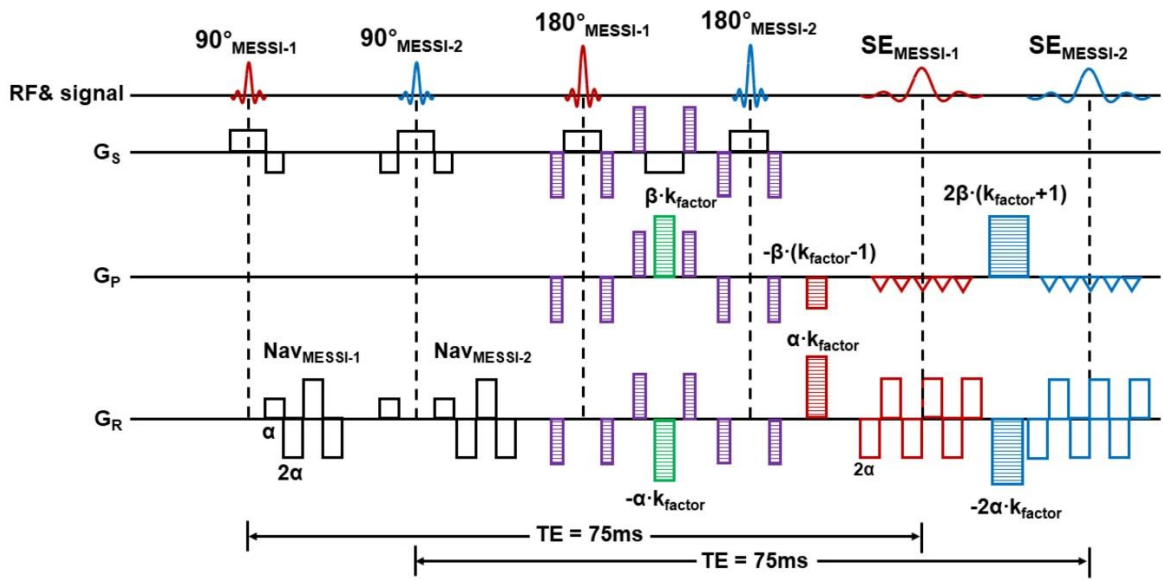

**Figure 3.**

**(A) k$_{factor}$=1**

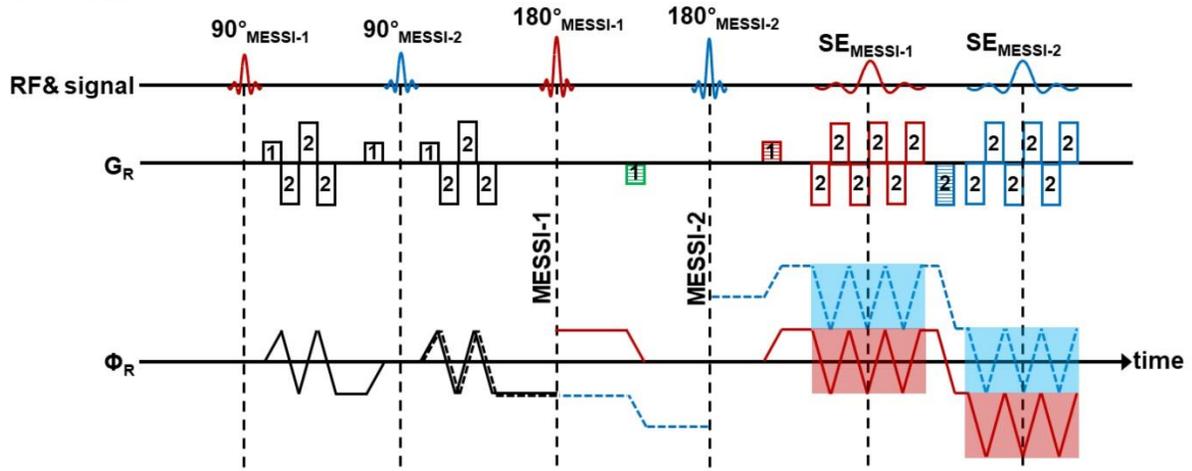

**(B) k$_{factor}$=2**

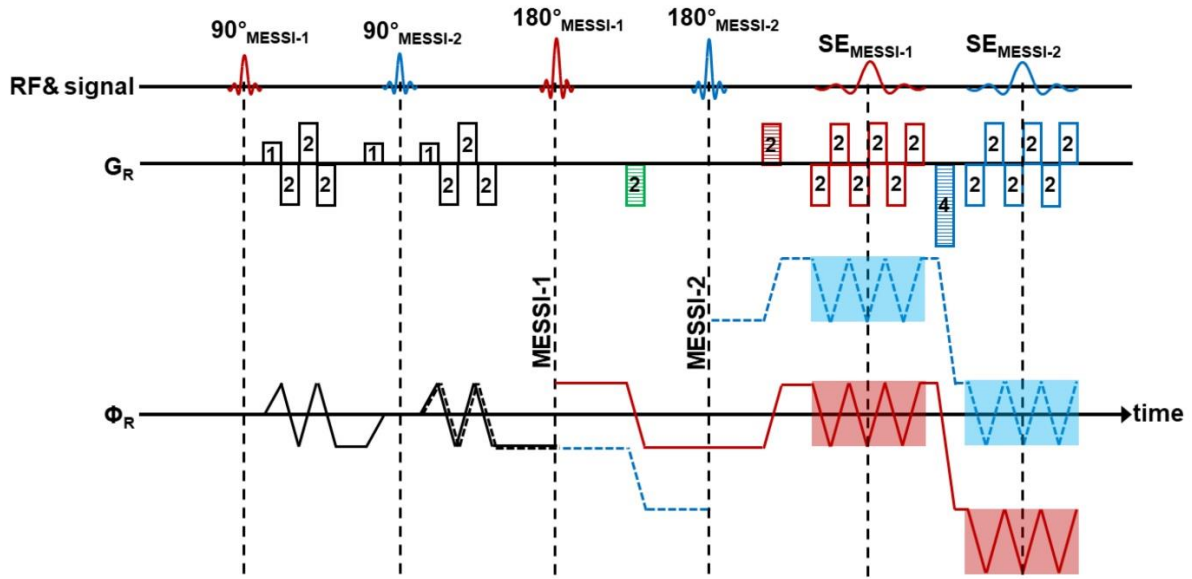

**Figure 4.**

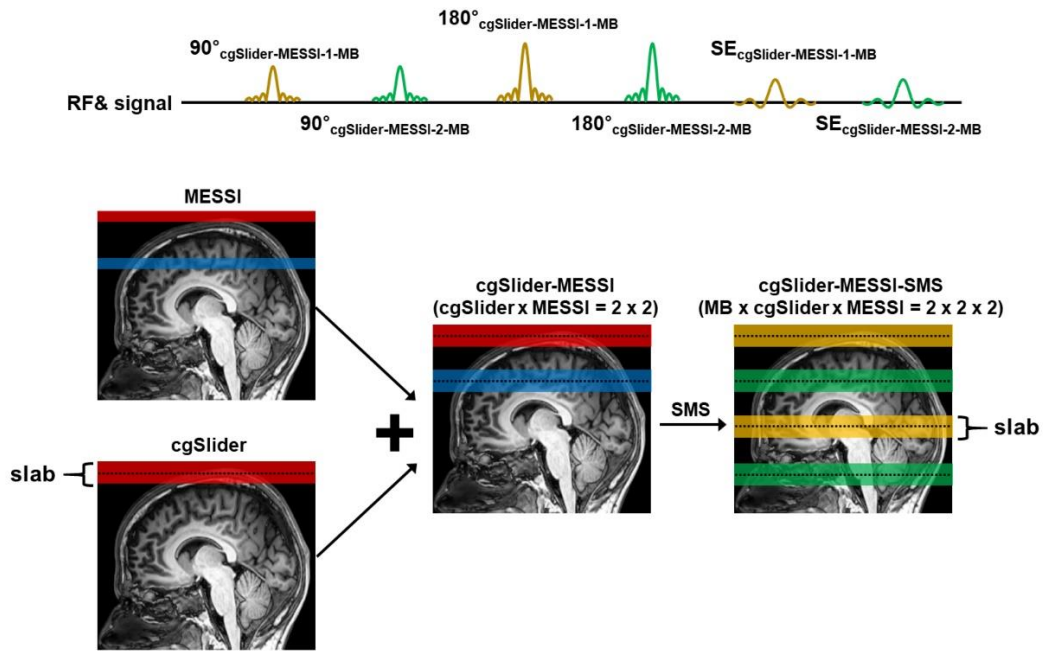

**Figure 5.**

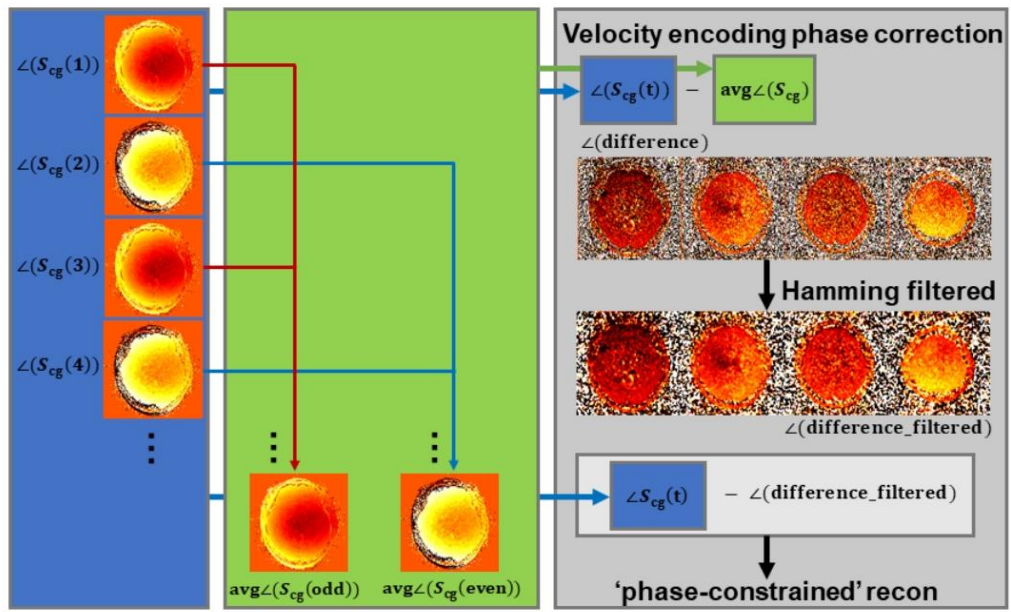

**Figure 6.**

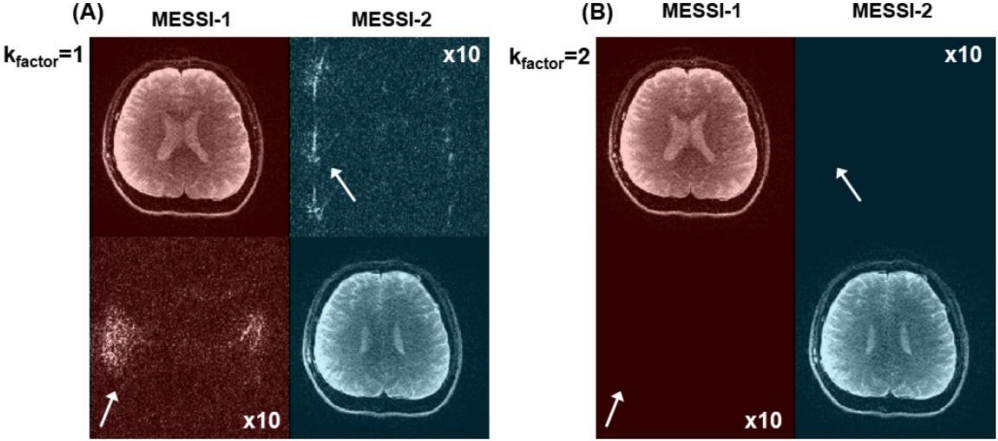

**Figure 7.**

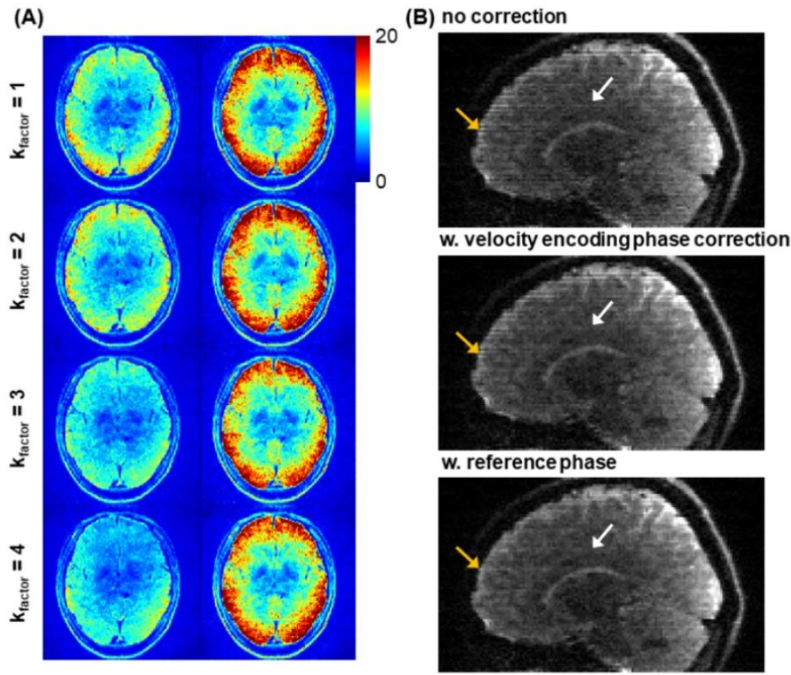

**Figure 8.**

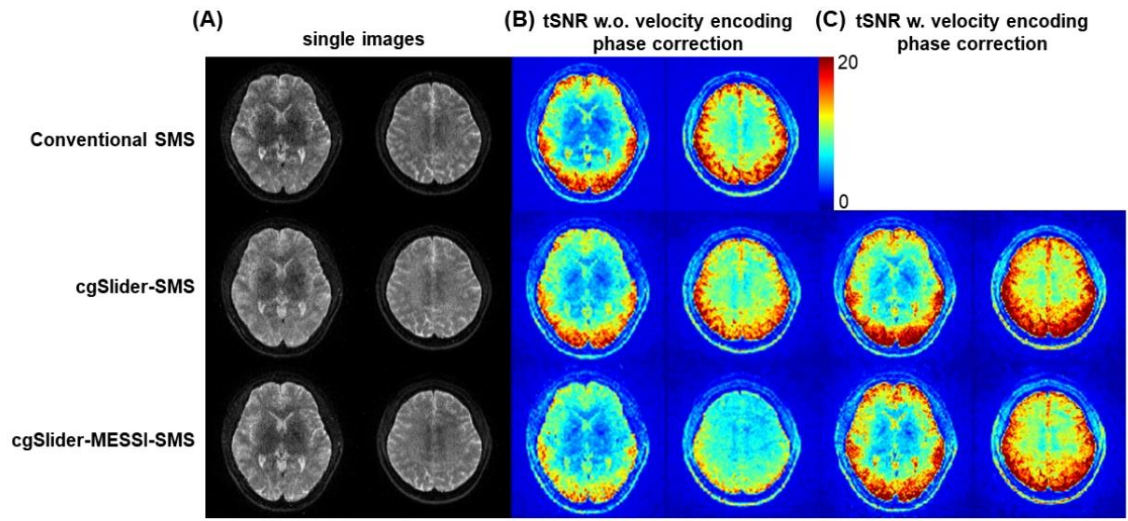

**Figure 9.**

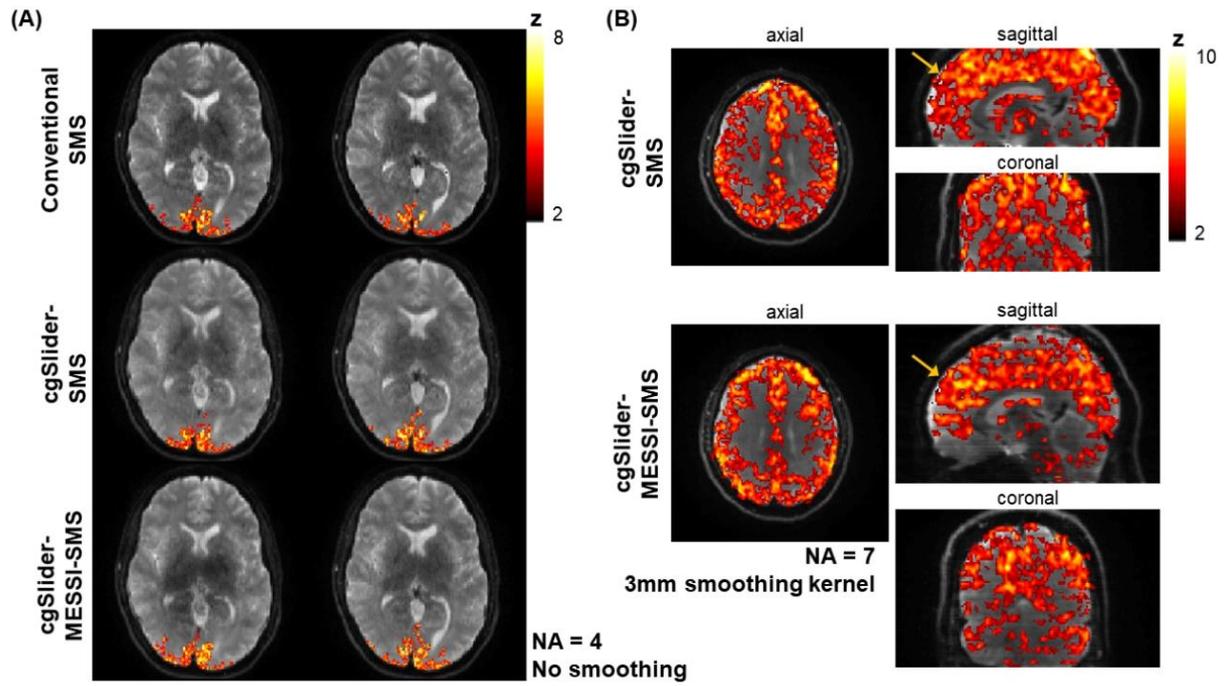

# Supporting Information

## Supporting background information

$S_{cg}$ is the cgSlider signal acquired at each time point ($n$). $S_A$ and $S_B$ are the magnitudes of sub-slices A and B, $\phi_A$ and $\phi_B$ are the corresponding image background phases of sub-slices A and B, respectively.

If $n$ is odd,

$$S_{cg}(n) = S_A(n)e^{j\phi_A(n)} + jS_B(n)e^{j\phi_B(n)}$$

$$S_{cg}(n)e^{-j\phi_A(n)} = S_A(n) + jS_B(n)e^{j(\phi_B(n)-\phi_A(n))}$$

$$S_{cg}(n)e^{-j\phi_A(n)} = S_A(n) + jS_B(n)(\cos\Delta\phi(n) + j\sin\Delta\phi(n)), where\ \Delta\phi(n) = \phi_B(n) - \phi_A(n)$$

$$S_{cg}(n)e^{-j\phi_A(n)} = S_A(n) + jS_B(n)\cos\Delta\phi(n) - S_B(n)\sin\Delta\phi(n)$$

$$Re(S_{cg}(n)e^{-j\phi_A(n)}) = S_A(n) - S_B(n)\sin\Delta\phi(n)$$
$$Im(S_{cg}(n)e^{-j\phi_A(n)}) = S_B(n)\cos\Delta\phi(n)$$

$$\begin{bmatrix} Re(S_{cg}(n)e^{-j\phi_A(n)}) \\ Im(S_{cg}(n)e^{-j\phi_A(n)}) \end{bmatrix} = \begin{bmatrix} 1 & -\sin\Delta\phi(n) \\ 0 & \cos\Delta\phi(n) \end{bmatrix}\begin{bmatrix} S_A(n) \\ S_B(n) \end{bmatrix} \quad (4)$$

If $n$ is even,

$$S_{cg}(n) = S_A(n)e^{j\phi_A(n)} - jS_B(n)e^{j\phi_B(n)}$$

$$S_{cg}(n)e^{-j\phi_A(n)} = S_A(n) - jS_B(n)e^{j(\phi_B(n)-\phi_A(n))}$$

$$S_{cg}(n)e^{-j\phi_A(n)} = S_A(n) - jS_B(n)(\cos\Delta\phi(n) + j\sin\Delta\phi(n)), where\ \Delta\phi(n) = \phi_B(n) - \phi_A(n)$$

$$S_{cg}(n)e^{-j\phi_A(n)} = S_A(n) - jS_B(n)\cos\Delta\phi(n) + S_B(n)\sin\Delta\phi(n)$$

$$Re(S_{cg}(n)e^{-j\phi_A(n)}) = S_A(n) + S_B(n)\sin\Delta\phi(n)$$
$$Im(S_{cg}(n)e^{-j\phi_A(n)}) = -S_B(n)\cos\Delta\phi(n)$$

$$\begin{bmatrix} Re(S_{cg}(n)e^{-j\phi_A(n)}) \\ Im(S_{cg}(n)e^{-j\phi_A(n)}) \end{bmatrix} = \begin{bmatrix} 1 & \sin\Delta\phi(n) \\ 0 & -\cos\Delta\phi(n) \end{bmatrix}\begin{bmatrix} S_A(n) \\ S_B(n) \end{bmatrix} \quad (5)$$

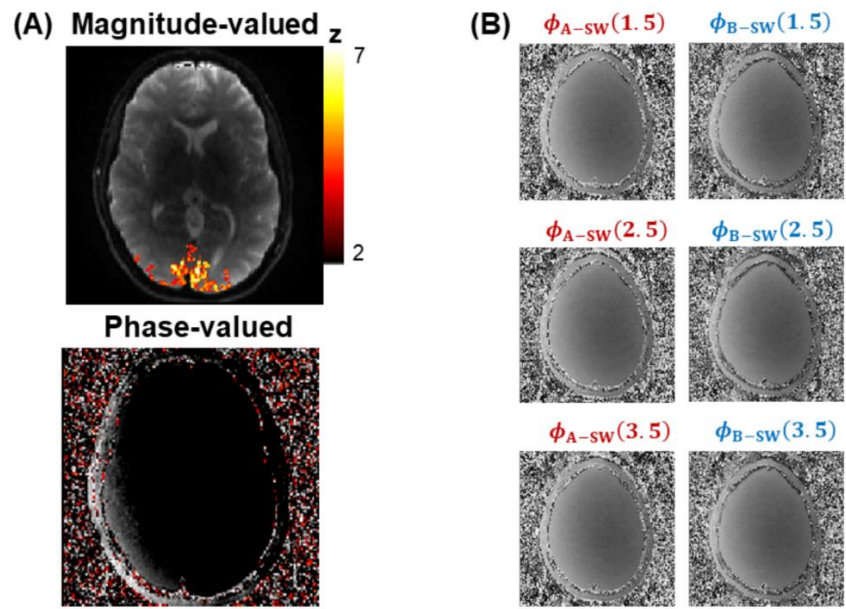

**Supporting Information Figure S1. (A)** Z-statistical maps obtained from signal intensity (top) and phase data (bottom) from a visual stimulation task. Detected fMRI responses are found only in the signal magnitude reconstruction with little/no fMRI responses detected in the phase image. **(B)** Estimated background phase from cgSlider acquisition and sliding-window reconstruction for two adjacent slices (A, B sub-slices depicted in Fig.1.) and three time points (1.5, 2.5, 3.5) with little phase variation over time and between sub-slices.

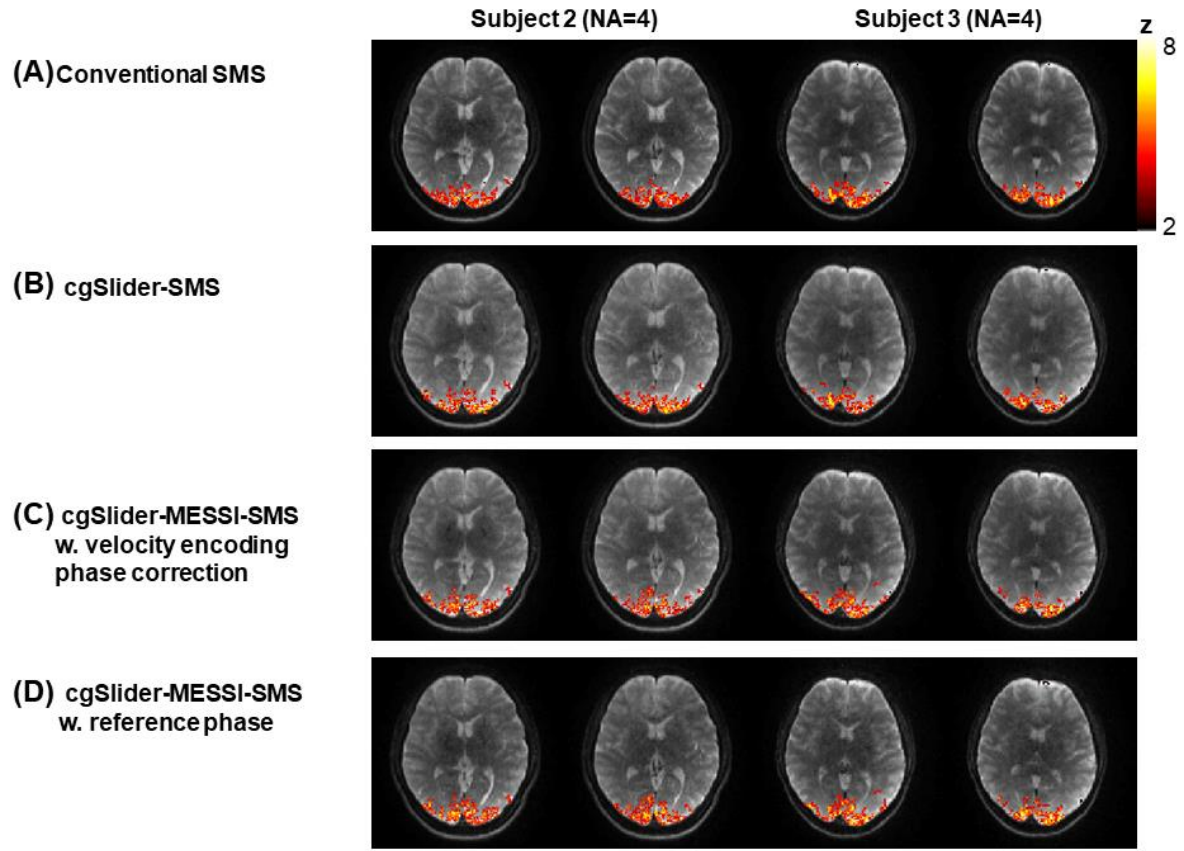

**Supporting Information Figure S2.** Z-statistic maps from visual stimulation with **(A)** conventional SMS, **(B)** cgSlider-SMS, **(C)** cgSlider-MESSI-SMS after velocity encoding phase correction, and **(D)** cgSlider-MESSI-SMS after correction using the reference phase. Intensity correction was performed for all images.

**Supporting Information Table S1.** Mean and standard deviation of z values (z > 2), and the number of activated voxels (clusters of minimum size of 30 voxels) from three subjects.

| Subject 1 | mean ± std z values | # of activated voxels |
|---|---|---|
| conventional SMS | 3.6 ± 1.2 | 1571 |
| cgSlider-SMS | 3.6 ± 1.3 | 1733 |
| cgSlider-MESSI-SMS | 3.5 ± 1.2 | 1591 |

| Subject 2 | mean ± std z values | # of activated voxels |
|---|---|---|
| conventional SMS | 3.2 ± 0.9 | 1700 |
| cgSlider-SMS | 3.5 ± 1.1 | 1922 |
| cgSlider-MESSI-SMS | 3.3 ± 1.0 | 1631 |

| Subject 3 | mean ± std z values | # of activated voxels |
|---|---|---|
| conventional SMS | 3.5 ± 1.1 | 1666 |
| cgSlider-SMS | 3.4 ± 1.1 | 1813 |
| cgSlider-MESSI-SMS | 3.5 ± 1.1 | 1535 |